\documentclass[fleqn,12pt,twoside]{article}
\usepackage{espcrc1}
\input{epsf}

\usepackage{graphicx}
\usepackage[figuresright]{rotating}

\newcommand{\AmS}{{\protect\the\textfont2
  A\kern-.1667em\lower.5ex\hbox{M}\kern-.125emS}}
\newcommand{\Lam}{\Lambda(1405) }

\title{Structure of $\Lambda(1405)$ and chiral dynamics\footnote{The title of the talk in the conference is ``Chiral dynamics of the two $\Lambda(1405)$ states''. }}

\author{D.\ Jido\address[TUM]{Physik-Department, Technische
         Universit\"at M\"unchen, D-85747 Garching, Germany}%
        \thanks{Attendance at the conference was financially supported
        by the Alexander von Humboldt Foundation.},
        J.A.\ Oller\address[MU]{Departamento de F\'{\i}sica, 
        Universidad de Murcia, 30071 Murcia, Spain},
        E.\ Oset\address[IFIC]{Departamento~de~F\'{\i}sica~Te\'orica~and~IFIC,~Universidad~de~Valencia,~46071~Valencia,~Spain},
        A.\ Ramos\address[BUS]{Departament d'Estructura i Constituents 
        de la Mat\`eria, Universitat de Barcelona, Diagonal 647, 
        08028 Barcelona, Spain},
        U.-G.\ Mei\ss{}ner\address[BUG]{HISKP, University of Bonn, 
        Nu{\ss}allee 14-16, D-53115 Bonn, Germany,\\
        FZ J\"ulich, Institut f\"ur Kernphysik (Theorie), D-52425
        J\"ulich, Germany}}

\begin{document}

\maketitle

\begin{abstract}
We report on a recent theoretical work on the structure of the $\Lam$ resonance within a chiral unitary approach, in which the resonance is dynamically generated in meson-baryon scattering.  Studying the analytic structure of the scattering amplitude, we have found that there are two poles lying around energies of $\Lam$ with different widths and couplings to the meson-baryon states. We discuss reactions to conform the double pole structure in experiment and elastic $K^{-}p$ scattering at low energies.  
\end{abstract}

\section{Introduction}
The $\Lam$ resonance has been a longstanding example of a quasi-bound state of meson and baryon appearing naturally in scattering theory with coupled channels~\cite{dalitz}. Modern formulations combining the chiral effective theory with unitary frameworks all lead to the dynamical generation of this resonance in the meson-baryon scattering~\cite{ChULam,oller,bennhold}. 
Yet, it was shown that in
some models, such as the cloudy bag model~\cite{fink}, one could obtain two poles close to the nominal $\Lambda(1405)$ resonance.
Also, in the investigation of the poles of the scattering matrix within the
context of chiral dynamics~\cite{oller}, it was found that there were two poles close to the nominal $\Lambda(1405)$ resonance both contributing to the $\pi\Sigma$ invariant mass distribution.  This was also the case in other works~\cite{twolam}.
Since the $\Lam$ lies below the threshold of $\bar KN$, the investigation of the $\Lam$ structure is  important also in view of low energy dynamics of anti-kaon and nucleon. 

\section{$\Lam$ in chiral unitary approach}

\begin{figure}[tbh]
\begin{minipage}[t]{75mm}
\epsfxsize=55mm
\centerline{\epsfbox{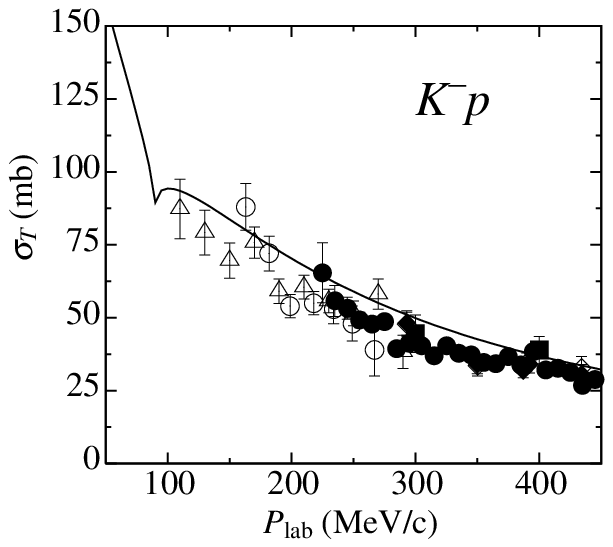}}
\caption{The total cross section of the $K^{-}p$ elastic scattering calculated by the chiral unitary approach. The cross sections of the other channels and the resources of the experimental data are shown in Ref.~\cite{bennhold}.
\label{fig:kpscatt}}
\end{minipage}
\hspace{\fill}
\begin{minipage}[t]{80mm}
\epsfxsize=70mm
\centerline{\epsfbox{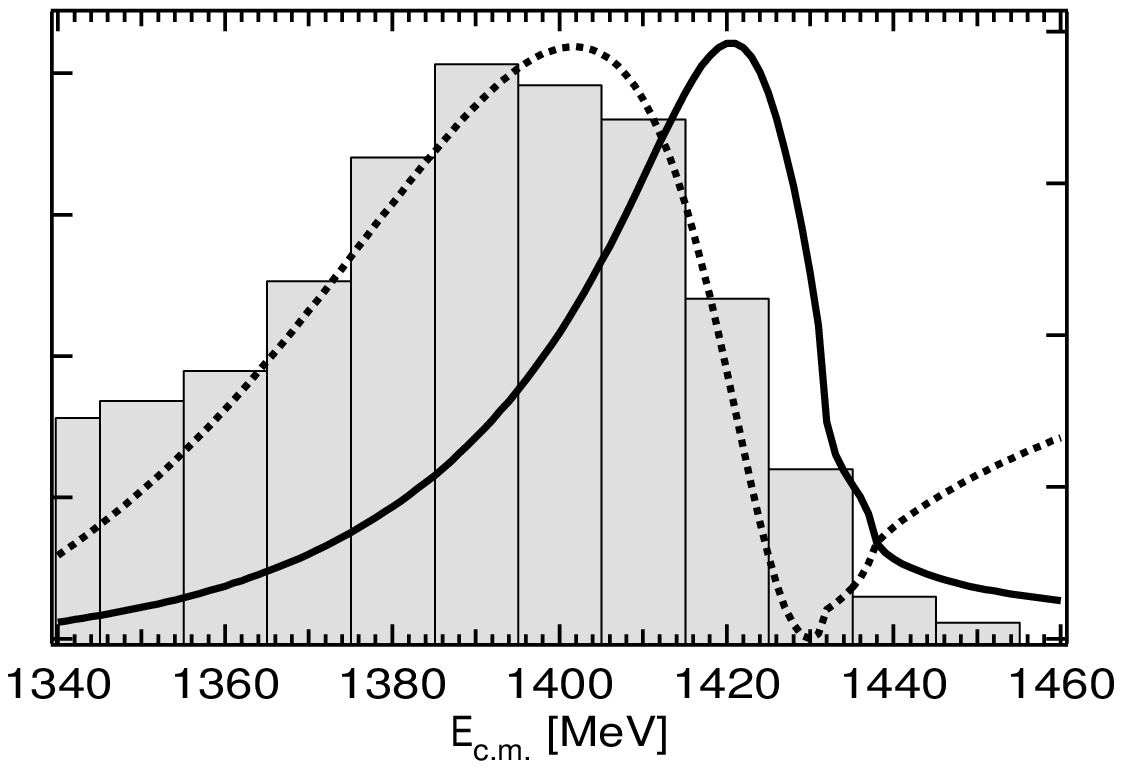}}
\caption{The $\pi\Sigma$ mass distributions with $I=0$ constructed from
  the $\pi\Sigma \rightarrow \pi\Sigma$ (dotted line) and $\bar{K}N\rightarrow
\pi\Sigma$ (solid line) amplitudes obtained by the chiral unitary approach~\cite{Jido:2003cb}. 
The histogram shows experimental
data~\cite{hist}.  Units are arbitrary.
\label{fig:massdist}}
\end{minipage}
\end{figure}

The $\Lam$ resonance is described here as a dynamically generated object in coupled-channel meson-baryon scattering with $S=-1$ and $I=0$
within the chiral unitary approach~\cite{lutz}.
We start with the scattering amplitude of the lowest order of the chiral perturbation theory, which is the Weinberg-Tomozawa term in case of the $s$ wave resonance. 
Respecting the flavor SU(3) symmetry, we consider all of the octet mesons 
($\pi$, $K$, $\eta$) and the octet baryons ($N$, $\Lambda$, $\Sigma$, $\Xi$) in the scattering channels. 
The unitary condition is imposed  by summing up a series of relevant diagrams non-perturbatively.
At this stage, the regularization of the loop integral brings parameters, which are the subtraction constants in the dispersion relation of the N/D method.
In the present work, the parameters are fitted so as to reproduce the threshold branching rations of $K^{-}p$ to $\pi\Sigma$ and $\pi\Lambda$ channels. Consequently, we obtain good agreements in the total cross sections of the $K^{-}p$ scatterings to several channels~\cite{oller,bennhold}. For instance, as shown in Fig.~\ref{fig:kpscatt}, the total cross section of the $K^{-}p$ elastic scattering is reproduced well in a wide region of the $K^{-}$ momenta. 
The details of the model are given in Refs.~\cite{oller,bennhold,Jido:2003cb}.

The good advantage of our approach is to obtain an analytic solution of the scattering equation under a low energy approximation in which one takes only the $s$-channel unitarity and limits the model space of the unitary integral to one meson and one baryon states~\cite{oller}. 
Exploiting this advantage, we search resonance poles for the scattering amplitudes with $S=-1$ and $I=0$ in the second Riemann sheet, 
which are summarized in Table~\ref{tab:jido0} together with
the coupling strengths of the resonances to the meson-baryon states as the residues of the amplitude at the pole position.

We see that there are two poles around the $\Lam$ energies below the threshold of the $\bar KN$ channel. 
The two poles are located so closely that the corresponding resonances do not separately appear on the real axis. Therefore, what is seen in the mass spectrum around the $\Lam$ energies is a superposition of these two states, as shown in Fig.~\ref{fig:massdist} as the dotted line.
Another interesting point is that the resonances show different coupling strengths: the lower resonance strongly couples to the $\pi\Sigma$ state, while the higher pole dominantly couples to the $\bar KN$ state. 
This is very important in investigating the structure of the $\Lam$ resonance in experiment.

\begin{table}[bt]
\caption{Pole positions and couplings to $I=0$
physical states from Ref.~\cite{Jido:2003cb}.
 \label{tab:jido0}}
{
\begin{tabular}{ccccccc}
\hline
 $z_{R}$ & \multicolumn{2}{c}{$1390 - 66i$} &
\multicolumn{2}{c}{$1426 - 16i$} &
 \multicolumn{2}{c}{$1680 - 20i$}  \\
 & $g_i$ & $|g_i|$ & $g_i$ & $|g_i|$ & $g_i$ & $|g_i|$ \\
 \hline
 $\pi \Sigma$ & $-2.5+1.5i$ & 2.9 & $0.42+1.4i$ & 1.5 & $-0.003+0.27i$ &
 0.27 \\
 ${\bar K} N$ & $1.2-1.7i$ & 2.1 & $-2.5-0.94i$ & 2.7 & $0.30-0.71i$ &
 0.77 \\
 $\eta\Lambda$ & $0.01-0.77i$ & 0.77 & $-1.4-0.21i$ & 1.4 & $-1.1+0.12i$ &
 1.1 \\
 $K\Xi$ & $-0.45+0.41i$ & 0.61 & $0.11+0.33i$ & 0.35 & $3.4-0.14i$ &
 3.5 \\
 \hline
 \end{tabular}}
\end{table}

The existence of the two poles is strongly related to the flavor 
symmetry~\cite{Jido:2003cb}. The underlying SU(3) structure of the chiral Lagrangians
implies that 
a singlet and two octets of dynamically generated resonances should appear, 
to which the $\Lambda(1670)$ and the $\Sigma(1620)$ would 
belong~\cite{bennhold},
and that the two octets get degenerate in exact SU(3) symmetric case.
In the physical limit, the SU(3) breaking resolves the degeneracy of the
octets, and, as a consequence, one of them appears quite
close to the singlet around the $\Lam$ energies.

\section{Discussion}

The double pole structure of $\Lam$ found here should be confirmed by new experiments. The idea to observe evidence of the double pole structure is to see the channel dependence of the $\Lam$ mass spectrum.
In case of having two poles close to each other with different coupling strengths to the meson baryon states, as in the present case, the resonant shape should depend on the production channels, since both poles contribute to the mass spectrum with weights having the different channel dependences~\cite{Jido:2003cb,Hyodo:2003jw}.  This is seen in Fig.~\ref{fig:massdist}, in which we show the $\pi\Sigma$ mass distributions with $I=0$ initiated  by the $\pi\Sigma$ (dotted line) and $\bar KN$ (solid line) channels. Thanks to the large weight of the second pole lying at higher energy with narrower width, the $\bar KN \rightarrow \pi\Sigma$ channel shows a quite different spectrum from the original one observed in the $\pi\Sigma \rightarrow \pi \Sigma$ channel. 

One problem here is that one cannot access the second pole directly from the $\bar KN$ channel, since the resonances lie below the threshold of the  $\bar KN$ state.
This leads us to consider indirect reactions for the $\Lam$ production initiated by $\bar KN$.
One possible candidate of the indirect reaction is the $K^{-} p \rightarrow \gamma \Lambda(1405)$ reaction~\cite{Nacher:1999ni}, in which the $K^{-}$ loses some energy by emitting a photon before the creation of the resonances. 
Another possibility is a photo-induced $K^{*}$ production, $\gamma p \rightarrow K^{*} \Lam$, provided in Ref.~\cite{Hyodo:2004vt}.
In this reaction, $K^{-}$ can be dominated at low energy in the $t$-channel exchange by use of a linearly polarized
photon and detection of angular dependence of the final pion and kaon. 
It has been found that this process is suitable to isolate the second resonance.
If one could confirm the double pole structure, it 
would be one of the strong indications that the $\Lam$ structure is
largely dominated by a quasibound meson-baryon component. 

Finally let us briefly comment on model dependence of the double pole structure. 
It has been recently pointed out by updated analyses of low-energy $K^{-}$-proton interactions~\cite{Meissner:2004jr,Borasoy:2004kk} that there is inconsistency between the elastic $K^{-}p$ total cross section and the new DEAR measurements for the energy shift and width of kaonic hydrogen. In the later work~\cite{Borasoy:2004kk}, on discussion of the $\Lam$ mass spectrum, they have used the next-to-leading order chiral Lagrangian in a coupled channel approach with parameters fitted paying strong attention to the DEAR measurements. Since the DEAR data suggests a smaller imaginary part of the $K^{-}p$ scattering length, their fit with the DEAR data underestimates the total cross section of the elastic $K^{-}p$ scattering and, consequently, the double pole structure is not so clearly seen. Our analysis, on the other hand, is supported by the perfect reproduction of the $K^{-} p$ total cross section,
as well as the other works which obtain two poles around the $\Lam$ energies.
Thus, precise measurements of the elastic $K^{-}p$ scattering at low energies are very important also to understand the $\Lam$ structure.

\section{Conclusion}

The present investigation of the meson-baryon scattering within the chiral unitary approach suggests that two resonances are dynamically generated around energies of the nominal $\Lam$, as well as reproducing well the cross sections of the $K^{-}p$ to the various channels. Since the two resonances are located very close to each other, what one sees in experiments is a superposition of these two states. 
The existence of the two poles can be found out by performing different
experiments of the $\Lam$ creation initiated by the $\bar
K N$ state.
New measurements of the elastic $K^{-}p$ scattering around threshold energies also would figure out the structure of $\Lam$.

\end{document}